# Chemical Vapor Deposition of Nitrides by Carbon-free Brominated Precursors

*Stefano Leone\*, Teresa Duarte, Hanspeter Menner, Jannik Richter, Lutz Kirste, Sven Maegdefessel, Felix Hoffmann, Byeongchan So, and Ruediger Quay*

S. Leone, T. Duarte, H. Menner, J. Richter, L. Kirste, S. Maegdefessel, F. Hoffmann, B. So, R. Quay: Fraunhofer Institute for Applied Solid State Physics IAF, Tullastrasse 72, 79108 Freiburg, Germany
E-mail: stefano.leone@iaf.fraunhofer.de, leonestefano@hotmail.com

Dedicated to the memory of Dr. Giuseppe Abbondanza, a pioneer of epitaxial growth in Italy, an enthusiast and brilliant scientist, a fantastic leader, and a humble person.

**Abstract**

The epitaxial growth of group 13-nitride semiconductors (GaN, AlN, and AlGaN alloys) for the mass production and fabrication of high-frequency and high-power devices relies on metal-organic chemical vapor deposition (MOCVD) using metal-organic molecules, also called precursors. While this growth method ensures high productivity and low operation costs compared to other methods, its most significant disadvantage lies in the presence of carbon atoms in the precursors, which are unavoidably incorporated into the epitaxial layers and hamper the performance of most types of fabricated devices. Carbon-free precursors for the CVD process could enhance the performance of high-frequency and high-power nitride-based devices while maintaining growth capability in industrial equipment. In this work, we implement gallium- and aluminum-brominated precursors, which contain no carbon atoms, to grow GaN and AlN layers in an industrial CVD system. We compare the results of this alternative CVD process with the conventional method using trimethyl precursors through several characterization techniques, indicating a clear reduction in optically active carbon-related defects.

# 1. Introduction

The decarbonization of our digital- and electronics-driven society relies on more energy-efficient and performing devices. Wide-band-gap semiconductors, such as AlN, GaN, and the AlGaN alloys (also known as group 13-nitrides), enable reliable and efficient electronic applications thanks to their physical properties. Devices based on these nitrides require complex heterostructures made of several thin layers with well-defined composition, abrupt interfaces, and precise doping, which are typically obtained by growing them on multi-wafers metal-organic chemical vapor deposition reactors (MOCVD). This manufacturing process involves transforming metal-organic compounds, composed of a metal and a carbon-containing ligand, into inorganic layers such as GaN, AlGaN, and AlN. The presence of the organic ligand in the precursor makes it impossible to prevent the incorporation of carbon in the inorganic epitaxial layer. Unfortunately, the presence of carbon often introduces undesired effects, such as unintentional doping, self-compensation, or impurity scattering, which limit the performance of these wide-band-gap semiconductors. To achieve the maximum performance of nitride semiconductors in a carbon-free world (i.e., with reduced $CO_2$ emissions), a carbon-free chemistry for the epitaxial growth of nitrides is required. We need a substantial improvement in the growth of these materials by transitioning from MOCVD to pure chemical vapor deposition (CVD) and adopting carbon-free precursors.

There are numerous issues related to the unintentional doping of AlN, GaN, and $Al_xGa_{1-x}N$ (with $0 < x < 1$) layers with carbon, as reported in numerous scientific publications. Carbon in GaN often occupies nitrogen lattice sites ($C_N$), acting as a deep-level impurity (amphoteric, capable of behaving as a deep donor or acceptor).[1, 2] Even at background levels ($10^{16} – 10^{17}$ at $cm^{-3}$), carbon can compensate intentional dopants, reduce free carrier concentration, and introduce trap states that degrade device performance, particularly under high-frequency and high-power operations. For example, high-voltage GaN power devices require lightly doped drift layers (~$10^{15}$ at $cm^{-3}$), and trace carbon can strongly influence doping control at these levels.[3] Carbon incorporation into the channel layer of AlGaN/GaN heterostructures is one of the factors contributing to the degradation of two-dimensional electron gas (2DEG) properties.[4] Carbon-related deep levels also degrade electron mobility and cause parasitic luminescence (e.g., blue and yellow/green bands) that indicate defects and traps.[5] Carbon acts as a compensating center also for p-GaN layers, reducing the effectiveness of magnesium (Mg) p-type doping; this leads to higher resistivity in the p-GaN layer and degraded gate control.[6] In AlGaN, the challenge is exacerbated: carbon impurities act as compensating acceptors, further hindering n-type doping, in addition to the intrinsic difficulty of activating dopants in high-Al materials.[7] Eliminating carbon in pure AlN is essential because carbon acts as a deep defect (typically substituting on N-sites) that absorbs deep-UV light (~4.7 eV, ~265 nm)[8] and can compensate intentional dopants. Avoiding carbon contamination from the precursors is therefore crucial for achieving high-purity epitaxial layers with predictable doping and superior device performance. In summary, the decarbonization of group 13-nitrides can bring several benefits from several technological aspects: better doping control, especially facilitating p-doping, growth on nitrogen polar substrates with lower carbon incorporation, enhanced device reliability and achievable breakdown, improved current stability, and reduced dynamic on-resistance,[9, 10] to mention the most important benefits.

Before diving into the growth of group 13-nitride using the most suitable methods, it is worth noting that the most used semiconductor - silicon, and the most used wide band-gap semiconductor - silicon carbide (SiC), are both grown by chemical vapor deposition (CVD), not by MOCVD. While silicon epitaxial growth is typically performed using chlorosilanes (mainly $SiHCl_3$ or $SiCl_4$),[11] SiC is grown epitaxially using either the standard chemistry $SiH_4$ + $C_2H_4$ or the advanced chlorinated-chemistry where one or both precursors are chlorinated (e.g., $SiHCl_3$, $CH_3Cl$, $CH_3SiCl_3$).[12-19] Other halogenated species can also be used[20] for the growth of Si and SiC. However, there is no need because the efficiency of chlorinated precursors and the established robust hardware of CVD reactors make their adoption the standard for mass production in the industry.

Surprisingly, to date, there are only a few scientific studies[21] describing attempts to grow nitrides in commercial MOCVD reactors using C-free precursors. Instead, the standard epitaxial growth of Al- and Ga-nitrides via MOCVD is purely based on metal-organic precursors: trimethylgallium (TMGa, $(CH_3)_3Ga$), trimethylaluminum (TMAl, $(CH_3)_3Al$), and triethylgallium (TEGa, $(CH_3CH_2)_3Ga$). The consequence of using metal-organic precursors is that the carbon atoms present in the precursors incorporate in the solid epitaxial layers, especially at those growth conditions that favor incomplete pyrolysis of the precursors – for example, at lower temperatures or low $NH_3$ flows (low V/III ratio), or at extremely high growth rates. Carbon is acknowledged to be a problem, especially when low growth temperatures (e.g., growth of multi-quantum wells, regrowth of ohmic contacts, growth of p-GaN) or high growth rates (e.g., for thick drift layers) are needed, unless it is intentionally incorporated via auto-doping[22] or extrinsic doping.[23] Raising the substrate temperature improves the decomposition of methyl/ethyl groups, reducing residual carbon. Still, there is a trade-off since GaN decomposes at high temperatures and, in the case of AlGaN alloys, those high temperature conditions can exacerbate the formation of compensating vacancy–dopant complexes.[2] Thus,

AlGaN growth must balance temperature to minimize carbon without inducing other defects. There are ongoing attempts to reduce C-incorporation by improving $NH_3$ cracking either with the aid of a plasma source,[24] or by laser-assisted growth;[10] otherwise by keeping the growth rate very low (e.g., by pulsed growth) or by modifying reaction chamber design to allow enormous $NH_3$ flows, since it is known that high V/III ratios help to minimize C-unintentional auto-doping.[25]

Remarkably, we demonstrate for the first time, to the best of our knowledge, the epitaxial growth of GaN and AlN in a commercial MOCVD reactor with carbon-free precursors. We report here on the epitaxial growth of GaN using $GaBr_3$ and the growth of AlN using $AlBr_3$, in combination with $NH_3$ as the nitrogen source, and open the possibility also of growing AlGaN alloys in a multi-wafer commercial reactor.

## 2. Growth process description

Up to date, the only way to grow epitaxially Al- and Ga-nitrides with C-free precursors is either by molecular beam epitaxy (MBE) or halide chemical vapor deposition (HCVD, also called halide vapor phase epitaxy - HVPE). However, both MBE and HCVD are excellent methods, but they have some limitations when it comes to manufacturing multi-wafers with large diameters and controlling complex multi-layer heterostructures with different doping and alloy compositions, respectively. Maintaining high productivity and uniformity is only possible with (MO)CVD reactors. MBE is an epitaxial growth method that provides an inherently C-free environment for the growth of group 13-nitrides. The precursors are solid metals sublimated to produce metal atom beams, while nitrogen is supplied either as $NH_3$ or as plasma-activated nitrogen, rendering the system C-free. Nevertheless, the excellent purity and control enabled by MBE[26] are counterbalanced by a lower growth rate and scalability difficulties, which prevent MBE from being adopted as a mass production method by tier 1 nitride semiconductor manufacturers. In HVPE, the precursors are typically gaseous hydrides and halides: gallium monochloride (GaCl) is generated in situ by reacting Ga metal with HCl gas, and it then reacts with $NH_3$ at the substrate to form GaN (with HCl as a byproduct). In the case of Al-nitrides, Al metal reacts with HCl to form $AlCl_3$. An inert carrier gas then transports this compound into the growth zone, where it meets $NH_3$, generating AlN on the substrate. Clearly, no organometallic compounds are used, so the process chemistry introduces zero intentional carbon.[1, 27] Besides being a C-free process, the key advantage is the very high growth rate achievable (above 100 μm h$^{-1}$) and the high layer thicknesses that can be grown. Nevertheless, impurities represent a common issue due to the high corrosiveness of the chlorinated environment. Quartz-free reaction chambers and BN-coated graphite components seem to attenuate this problem [5], resulting in low oxygen and silicon impurities. However, while the design of HVPE reactor favors the deposition of very thick layer to obtain ingots to be cut in several substrates, they are not optimized for the deposition of highly-uniform thin heterostructures with multi-doped layers and precise alloy control, besides they are limited in reproducibility, as required by the semiconductor industry for devices fabrication, although a few reports have addressed this issue to some extent.[28] An implementation of chlorinated chemistry, as used in HVPE, to MOCVD of multi-wafers reactors has not been successful so far, mainly due to the high corrosiveness of chlorinated species. These species etch quartz and SiC-coated components and also attack most metals, which MOCVD reactors are made of. As an example, $GaCl_3$ would require Hastelloy bubblers because of its high corrosiveness even for stainless steel. Besides, the low vapor pressure of the Cl-based precursors requires heating of the bubbler from 100 °C (for $GaCl_3$) to above 150 °C (for $AlCl_3$). The best approach would be to employ a less aggressive halogen, such as bromine. Indeed, brominated and iodinated Ga- and Al-compounds are generally less aggressive than their fluorinated and chlorinated analogs, due to their lower hydrolysis potential and the lower Lewis acidity compared to more electronegative halides. In the seventies, Chu *et al.* reported the growth of GaN via a CVD-like process using the $GaBr_3 \cdot 4NH_3$ complex as gallium and nitrogen source.[21] This complex has a very low vapor pressure, which requires heating above 200 °C to achieve sufficient vaporization. This limitation restricts process control and imposes a high thermal load on the source. Besides, the precursor is thermally unstable and tends to decompose upon heating, inaccurately releasing $NH_3$. Since this molecule is highly hygroscopic, it must be placed directly inside the reaction chamber, similar to how sources are handled in HCVD reactors. Since these experiments were reported, there has been barely any report that describes a successful reproduction of those results. On top of this, another argument against the adoption of brominated chemistry came from thermodynamic analysis that showed how brominated group 13-compounds have a chemical potential that is not suitable for achieving a controlled composition of group 13-nitride alloys.[29]

To summarize: the extreme corrosiveness of chlorinated compounds, the successful but at extreme conditions and challenging to reproduce experiments with the $GaBr_3 \cdot 4NH_3$ complex, and eventually the thermodynamic calculations predicting the impossibility of growing (Al, Ga)N alloys seem to point to the impossibility of depositing GaN, AlN, and AlGaN alloys with a C-free precursors CVD process.

The feasibility of growing group 13-nitrides with halogenated precursors in MOCVD reactors begins with understanding the nature of the chemical bonds formed between group 17 (halogens) elements and Ga or Al, identifying suitable precursor candidates and their main physical properties, and ultimately assessing their vaporization behavior in a commercial (MO)CVD reactor. Obviously, the reactor has to be capable of handling such species, avoiding highly corrosive compounds that would end up in corroding reactor components and contaminating the epitaxial layer, while at the same time be able to be generate a molar flow of at least one µmol min$^{-1}$ that can be controlled by digital controllers (e.g., mass flow controller or pressure controller/gauge). Finally, the growth of GaN and AlN epitaxial layers by using solely these halogenated precursors and NH$_3$ would validate the feasibility of the process.

## 2.1. Selection of precursors

Fluorinated and chlorinated group 13 precursors are known to be highly aggressive even towards metal components (stainless steel) and quartz. This explains why the trend in HCVD reactors is to generate a chlorinated source directly inside the reaction chamber, rather than in a standard stainless-steel bubbler, and minimize the use of quartz components, as in the quartz-free HCVD reactors reported elsewhere [5]. Iodine (I) and astatine (At) or even tennessine (Ts) (the last two being radioactive and not even occurring naturally) have much lower electronegativity, and the molecules formed with Al or Ga are expected to be less aggressive than the fluorinated and chlorinated ones. However, the physical properties of species such as AlI$_3$ or GaI$_3$ indicate very high melting points (192 °C and 212 °C, respectively) and, consequently, extremely low vapor pressure, making them not practicable for use in an MOCVD reactor. Instead, the brominated species may be the best compromise. While the monobromides (AlBr and GaBr) are very unstable and decompose before melting, the tribromides AlBr$_3$ and GaBr$_3$, are very stable and, already from their melting point temperatures (97.5 ˚C and 121.5 °C, respectively), they can be expected to have an acceptable vapor pressure to generate the desired molar flow above one µmol min$^{-1}$. Indeed, these compounds have a vapor pressure high enough to be operated in proper vaporizers for low vapor pressure species, such as the one we have developed and patented at Fraunhofer IAF[30] for the MOCVD of nitride epitaxial layers with Sc- and Y-precursors.[31-46] To find the appropriate temperature at which to keep the bubbler, we need to calculate the vapor pressure and the conditions (carrier gas and pressure) at which to vaporize the precursor. As reported in literature and inorganic chemistry textbooks,[47] the vapor pressure of AlBr$_3$ can be calculated with the formula (Antoine equation):[48]

$$\log_{10} P = 14.78 - \frac{4700}{T}$$

The vapor pressure of the chemical compound ($P$) is in Pascals, and the temperature in Kelvin. The vapor pressure is approximately 1.5 mbar at 100 °C. In the case of $GaBr_3$, we can use the following formula:[49]

$$\log_{10} P = 16.1 - \frac{5250}{T}$$

In this case, the vapor pressure at 100 °C is slightly lower, amounting to circa one mbar. Both precursors are stable up to 150 °C and do not tend to decompose below this temperature, making them suitable for a controllable and reproducible vaporization of the source for a CVD growth process. As a comparison, the standard metal-organic precursors for Al (TMAl) and Ga (TMGa) have a vapor pressure above 400 and 4000 mbar at 100 °C, respectively. They are typically operated at much lower temperatures and higher bubbler pressures, resulting in vapor pressures of 9.8 mbar (TMAl kept at 17 °C) and 91.25 mbar (TMGa kept at 0 °C). Based on the vapor pressure curves of the brominated precursors, we decided to maintain the $AlBr_3$ source at 110 °C and the $GaBr_3$ source at 135 °C, in both cases slightly above the melting temperatures to facilitate bubbling of the carrier gas and prevent channeling effects in a solid source, achieving vapor pressures of 3.26 and 17.26 mbar, respectively, which are quite similar to that of TMAl. The main physical characteristics of the standard and brominated precursors are listed in **Table 1**.

**Table 1.** Comparative Properties of TMGa, TMAl, $GaBr_3$, and $AlBr_3$

| Precursor | Melting Point (°C) | Use Temp (°C) | P_vap @ 100°C (mbar) | P_vap @ T_use (mbar) |
|---|---|---|---|---|
| TMGa | -15 | 0 | ~620 | 91.25 |
| TMAl | 15 | 17 | ~440 | 9.8 |
| $GaBr_3$ | 121.5 | 135 | ~1.0 | 17.26 |
| $AlBr_3$ | 97.5 | 110 | ~1.5 | 3.26 |

The molar flow can be calculated according to the general gas equation $PV = nRT$, where $P$ is the pressure (in Pascal), $V$ is the volume (in cubic meter), $n$ is the amount of moles, $R$ is the gas constant (8.314462 J K$^{-1}$mol$^{-1}$), and finally $T$ is the temperature (in Kelvin). We calculate the molar flow $n$ of each precursor in moles per minute, assuming $P$ to be the pressure in the bubbler of the precursor, $V$ the volume coming out of the bubbler upon flowing a carrier gas, and $T$ as the temperature at which the bubbler was heated. In our specific case, by keeping the AlBr$_3$ bubbler at 110 °C and ~300mbar, and a volume of 500 sccm coming out of the bubbler, we calculate that the resulting molar flow is above 240 µmol min$^{-1}$. In the case of GaBr$_3$, operating it at the same pressure and volume but heating the bubbler to 135 °C, we achieved a molar flow above 1000 µmol min$^{-1}$. These values are quite comparable to those obtained with TMGa and TMAl maintained at the previously indicated temperatures (0 and 17 ˚C), a bubbler pressure of 1300 mbar, and a gas volume of 50-100 sccm. The calculations show that the two brominated precursors of Al and Ga can be easily controlled and maintained at molar flows comparable to those of the standard metal-organic Al and Ga precursors, potentially enabling the CVD growth of GaN, AlN, and AlGaN alloys.

As a drawback, these brominated precursors are not commonly manufactured for the semiconductor industry. Therefore, no electronic-grade purity has been pursued so far by chemical manufacturers. Besides oxygen, a common contaminant is silicon, which comes from the use of glass or quartz containers for storing, synthesizing, purifying, and handling AlBr$_3$ and GaBr$_3$. Both species are powerful Lewis acids and very prone to react with silicon dioxide present in glass and quartz. Even the standard TMGa and TMAl precursors, which people began using last century for the MOCVD of group 13-As/P or nitrides, did not have high purity. Once market demand rose, manufacturers invested more effort in purification and other steps to produce electronic-grade materials suitable for the semiconductor industry. Likely, the same will happen once the market demand for AlBr$_3$ and GaBr$_3$ grows. Up to date, commercially available brominated Al and Ga precursors suffer from high Si-contaminations, in the order of a few ppm.

**2.2. CVD process setup**

A chemical vapor deposition process using brominated precursors could also be considered a halide-CVD process. However, in a typical HCVD/HVPE reactor, the sources are located inside the reaction chamber, and their molar flows cannot be controlled with the same accuracy as in a multi-wafer (MO)CVD reactor, where the sources are placed in bubblers and their flows are precisely controlled by means of mass flow controllers and pressure monitoring systems. Consequently, we refer to our process as a "CVD process".

In our experiments, we use a commercial Aixtron reactor with a close-coupled showerhead injection system equipped with three plenums, featuring a reaction chamber with a susceptor fitting 6x2" wafers and up to a 1x6" wafer. The reactor was baked before each growth run and coated with AlN after every reactor maintenance to minimize possible contaminations (e.g., Si,

C, and O) from the SiC coating or the few quartz parts of the reaction chamber. In our experiments, we used single sapphire (0001 $Al_2O_3$) 4" (100 mm) substrates. We used the same setup developed and patented for the vaporization of Sc and Y-precursors.[30] The gas lines upstream and downstream of the sources are heated and kept at a temperature of at least 10-15 °C above the bubbler temperature to prevent condensation in the lines. Typically, we monitor the carrier gas flow in the bubbler using a mass flow controller placed before the heated zone, and we monitor the pressure at the bubbler outlet with a pressure gauge capable of operating at high temperatures. The injection system of the reaction chamber is heated to 50-85 °C to prevent or reduce condensation in the showerhead plenums. Additionally, the exhaust gas lines are heated to 85 °C to minimize condensation before a large particle filter collects most of the by-products, such as $NH_4Br$. A wet scrubber collects and abates the exhaust gases. We used two different vaporizing systems for $GaBr_3$ and $AlBr_3$. In the case of $GaBr_3$, the bubbler was heated by a heat jacket, and the gas lines and valves by heat tapes. In the case of $AlBr_3$, as detailed in our patent, the bubbler is located inside an oven, and the runline is heated through a coaxial tube filled with a fluid regulated by a temperature controller. More details are described in our patent application about the CVD growth of (Al,Ga)N nitrides by brominated precursors.[50]

Similar growth conditions were adopted to test each precursor. We selected nitrogen as the carrier gas to minimize the formation of HBr as a byproduct, although HBr and $NH_4Br$ could still be formed from reactions between the bromide precursors and $NH_3$. A fixed growth temperature of 1200 °C and reactor pressure of 35 mbar were used for the experiments, while the $NH_3$ flow was set to 1100 sccm for the growth of AlN and 8000 sccm for the growth of GaN. At this initial stage of our research, the growth parameters and the growth rates were not optimized to a level that enabled the growth of thick epitaxial layers; consequently, we used AlN or GaN templates prepared with the standard MOCVD process for all the experiments. AlN templates, i.e., a 1 μm thick AlN deposited on c-plane sapphire, prepared on another MOCVD reactor by using TMAl, were used as a substrate for the growth of an AlN layer with $AlBr_3$, while in the case of the growth of GaN with $GaBr_3$, a similar template but with a 2 μm thick GaN layer grown using TMGa was used.

## 3. Results
### 3.1. Growth of GaN with $GaBr_3$

We deposited a 300 nm thick layer of GaN on the abovementioned GaN templates using a molar flow of $GaBr_3$ of about 180 μmol $min^{-1}$, achieving a growth rate of up to 0.4 μm $h^{-1}$. Attempts to increase the molar flow were not successful due to the setup used (bubbler heated with a heat jacket). As previously experienced with Sc- and Y-precursors, heating the bubbler with a heat jacket does not ensure proper heating, especially at the bubbler valves, resulting in the condensation of precursor in the gas line and bubbler valves. Consequently, after 10 to 15 minutes of flowing the precursor, the pressure in the gas line rose, and the flow must be interrupted to purge the lines. This effect was even more dramatic for higher molar flows. Future experiments with GaBr$_3$ will be conducted using a setup where the bubbler is placed inside a heated oven, along with a coaxially heated gas line, which typically does not exhibit this issue at all. The morphology of the grown layer was very smooth and crack-free, as shown in the figure obtained with atomic force microscopy (AFM) measurements in tapping mode with different scan sizes (**Figure 1**). The roughness RMS (root mean square) of a 2x2 µm$^2$ area was only 0.5 nm. The appearance of the surface morphology is surprisingly good, given the use of nitrogen as the carrier gas. We speculate that, thanks to the use of a high growth temperature, low pressure, and high NH$_3$ flow (which brings some hydrogen anyway to the gas phase), we have achieved a lateral 2D growth mode despite the absence of hydrogen as the carrier gas.

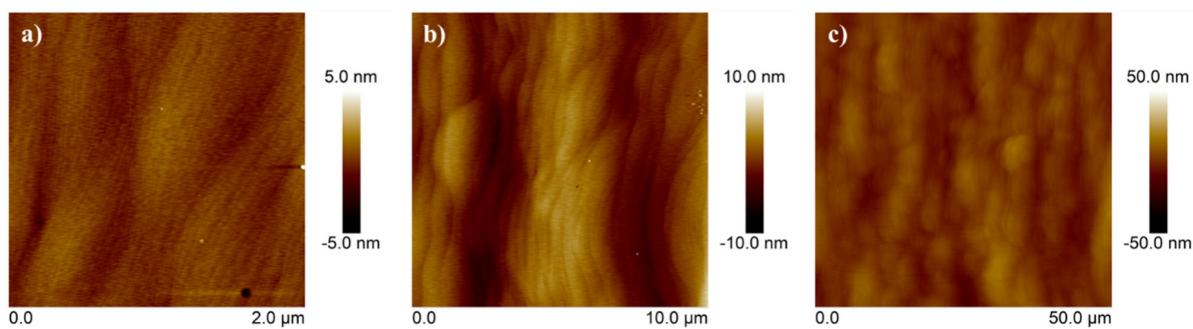

**Figure 1**: AFM analysis in tapping mode on a 300 nm thick GaN epitaxial layer deposited on GaN/Al$_2$O$_3$ templates with GaBr$_3$. The RMS is: a) 0.5 nm in a 2x2 µm$^2$ area, b) 2.1 nm in a 10x10 µm$^2$ area, c) 4.6 nm in a 50x50µm$^2$ area.

Further analysis were done and compared with a GaN layer that was grown using TMGa in the same MOCVD reactor and with the same growth conditions as those used with GaBr$_3$ (growth temperature of 1200 °C, NH$_3$ flow of 8000 sccm – corresponding to a V/III > 2000) except for using hydrogen as carrier gas and a reactor pressure of 200 mbar, rather than 35 mbar used for GaBr$_3$, which were needed to favor a higher molar flow of GaBr$_3$. We measured the sample with a Time of Flight secondary ions mass spectrometry tool (ToF-SIMS), but we could not appreciate any difference in the intensity of the carbon signal between the layer grown with GaBr$_3$ and the one grown with TMGa, including the template GaN layers. At this stage, the presence of carbon measured by SIMS is unclear. New measurements with a more sensitive SIMS tool are required and will be performed in the near future to reach a scientific conclusion.

A Br contamination can be excluded for the time being, as the measured concentration by ToF-SIMS was in the same order of magnitude as the one measured in the template used as substrate. Yet, a Si-contamination was noted, since Si at the ppm level was already in the source.

Optical analysis by cathodoluminescence (CL) and photoluminescence (PL) showed a clear distinction between the samples grown with TMGa and GaBr$_3$. Cathodoluminescence (CL) measurements were performed at an acceleration voltage of 5 kV, a beam current of 1.62 nA, scanning an area of at least 10 × 10 µm². The interaction volume (75% energy cut off) in depth was estimated to be 230 nm for AlN and 150 nm for GaN,[51-53] which is smaller than the thickness of the epitaxial layers studied in this work. CL spectra from 340 nm to 750 nm were recorded with a 300 grooves mm$^{-1}$ grating at RT. **Figure 2** shows the CL spectra of GaN layers grown with GaBr$_3$ (purple) or TMGa (blue). The spectra are normalized to near-band-edge (NBE) emission at 362 nm, which is related to exciton emission.[54] The blue luminescence (BL) and yellow luminescence (YL), centered at 430 nm and 560 nm, are defect-related.[55-56] There is evidence that they are related to isolated carbon defects or carbon-containing complexes.[57-60] The inset shows the ratio of CL intensities of the BL and YL. The peak at 724 nm is the second-order NBE emission.

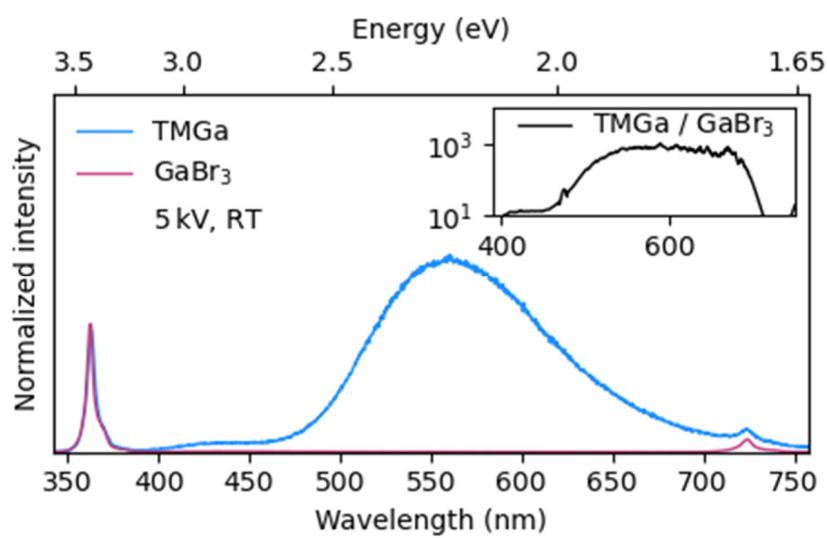

**Figure 2**: CL spectra from 340 nm to 750 nm of GaN layers grown with GaBr$_3$ (purple) and TMGa (blue), normalized to NBE (5 kV, 1.62 nA, RT). The GaBr$_3$-sample shows reduced emission intensities over the entire spectrum compared to the TMGa-sample. The inset shows the ratio of CL intensities of both samples.

The BL emission of the GaN layer grown with GaBr$_3$ is reduced by a factor of 10 and the YL by up to three orders of magnitude, compared to the one grown with TMGa. From the CL measurements, there is a reduction in all emission bands with respect to the free-exciton line, most notably the YL. In the case of the sample obtained with GaBr$_3$, the BL and YL are more than two and three orders of magnitude lower than the NBE, indicating very high optical purity of the GaN layer.

The PL measurements showed that the GaN sample grown with TMGa presents blue luminescence (~2.84 eV) at room temperature (RT) (**Figure 3**), which can be attributed to impurity-related defects, possibly a substitutional carbon ($C_N$) or its hydrogenated complex ($C_N$-$H_i$).[61-63] In the figure, the spectra are normalized to the NBE emission maximum of each sample, respectively. Both samples exhibit an NBE emission due to donor-bound exciton (DX) transitions. The sample grown with TMGa exhibits blue luminescence (BL), typically associated with carbon defects. In fact, under the optimized growth conditions using GaBr$_3$, this blue luminescence was not observed in either CL or PL measurements. It is important to note that PL and CL measurements cannot be used to assess the presence of carbon in the layers. However, the presented results may hint at the possibility of non-luminescent active defects.

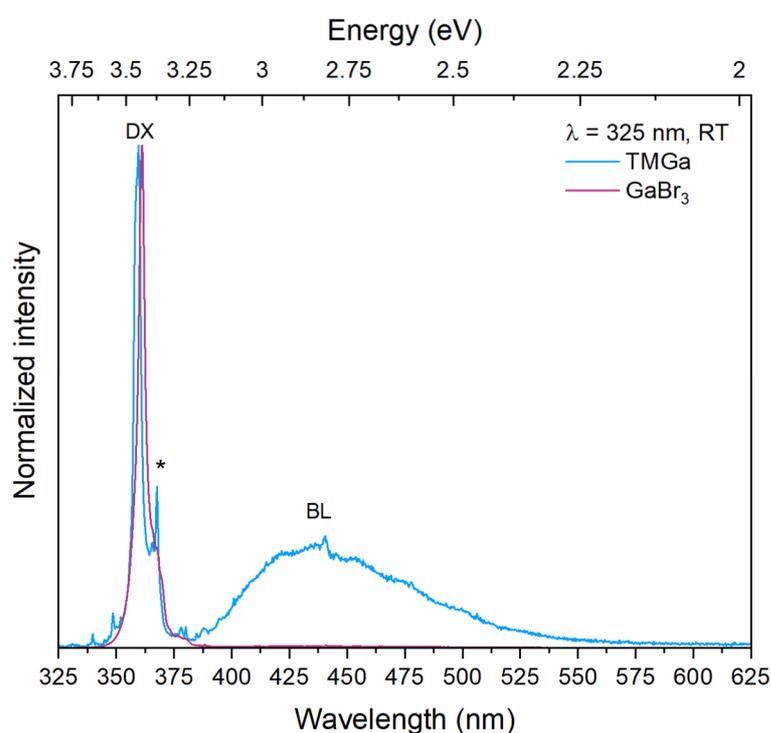

**Figure 3**: Room temperature (RT) PL spectra obtained with the excitation of a 325 nm laser line, acquired in the UV-Vis range, of a GaN layer grown with different precursors, GaBr$_3$ (purple) and TMGa (blue). * indicates a measurement artifact.

The full width at half maximum (FWHM) of ω-scans of the 300 nm thick GaN layer peak is influenced by the FWHM of the thick GaN epitaxial layer underneath,[64] as measured by high-resolution X-ray diffraction (HR-XRD). The FWHM for this GaN sample was less than 200 arcsec for the 0002 reflection, whereas it was less than 250 arcsec for the -1102 reflection, which is indicative that the crystal quality of the template was preserved (**Figure S1**). Finally, we performed electrical measurements of the sheet resistance using Eddy-current measurements and obtained the expected very high values of ~ 40 kΩ sq$^{-1}$. However, this value corresponds to the upper sensitivity limit of our eddy-current measurement setup, which cannot reliably quantify resistances above this threshold. Therefore, the high measured resistance should not be interpreted as evidence of strong compensation effects, but rather as an indication that the layer is highly resistive within the detection limit of the instrument. This observation is, nevertheless, consistent with the absence of blue and yellow luminescence in PL/CL, suggesting a low density of electrically and optically active carbon-related centers.

## 3.2. Growth of AlN with AlBr$_3$

We deposited AlN epitaxial layers with AlBr$_3$ (heated at 110 ˚C) on top of 1 μm thick AlN/Al$_2$O$_3$ templates at similar conditions as those used for GaN with GaBr$_3$ (heated at 135 ˚C). We have achieved growth rates up to 0.8 μm h$^{-1}$ at a molar flow close to 200 μmol min$^{-1}$. First experiments with a higher molar flow of 480 μmol min$^{-1}$ resulted in growth rates close to 1.8 μm h$^{-1}$, but this will be further tuned and discussed in future studies. In this case, crack-free and smooth surfaces were also obtained, although there is evidence of a 3D-like growth mode that needs to be optimized. The roughness of the layer measured by AFM was about 1.2 nm on a 2x2 μm$^2$ scan area (**Figure S2**). A sample with ~300 nm thick AlN deposited with AlBr$_3$ was analyzed by ToF-SIMS. A minimal decrease in carbon was observed compared to the layer deposited with TMAl, and new measurements with a more sensitive SIMS tool will be performed.

Optical analysis shed light on the improved quality of the layer deposited with AlBr$_3$. CL measurements were performed on AlN epitaxial layers grown with AlBr$_3$ or TMAl at similar growth conditions but with hydrogen as carrier gas (**Figure S3**). Both samples show a red luminescence band (RL, 600–850 nm) and a blue luminescence band (BL, 300–550 nm). The BL may be ascribed to vacancy–oxygen defect complexes.[65] The origin of RL has not yet been investigated. The AlBr$_3$-grown sample exhibits a reduced emission in BL and RL by at least a factor of 2.5, as seen in the inset of Figure S3. A carbon-related emission is reported at 265 nm,[66] attributed to substitutional carbon on the nitrogen lattice site.[8] In the CL spectra (inset

Figure S3), we observe a reduction in emission within this spectral range by up to one order of magnitude in the AlBr$_3$-grown sample.

The NBE shown in **Figure 4** consists of four contributions. We attribute the intense emission at 204.6 nm to the free-exciton emission, and the emission lines at 206 nm, 208 nm, and 212 nm to defect-bound excitons.[67-69] The bound-exciton emission at 208 nm (normalized to the free-exciton emission) is almost 20 times lower in the AlN layer deposited using AlBr$_3$ than in the one deposited with TMAl, which is a striking result indicating the high quality of the AlBr$_3$ grown layer. The sharp free exciton line, lower bound exciton emission, and reduced defect-band emission by at least a factor of two suggest a higher crystal quality and reduced defect density for the AlBr$_3$ sample. Carbon-bound excitons could explain the NBE spectra; however, a more detailed investigation of the exciton emission is needed.

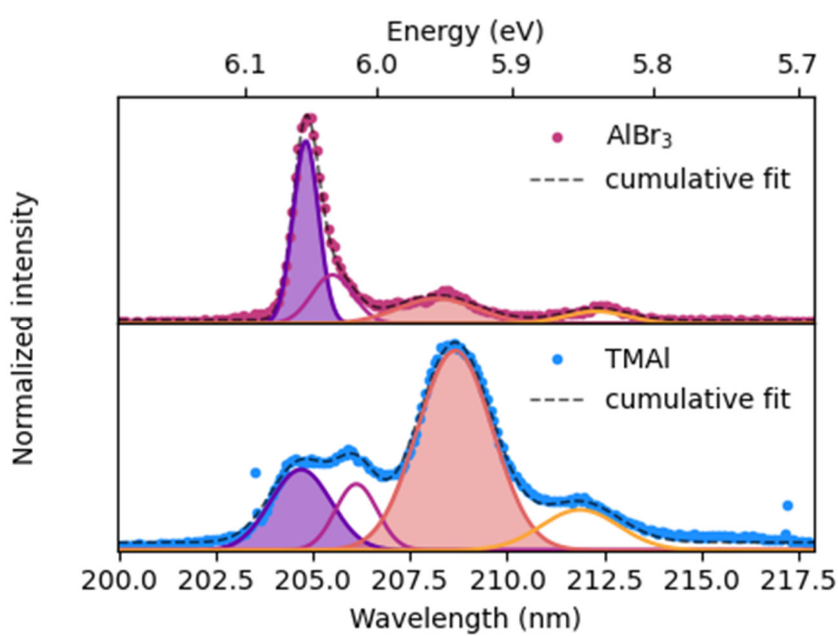

**Figure 4**: CL near-band-gap emission of AlN layers grown with AlBr$_3$ (upper) or TMAl (lower), normalized to the free-exciton emission (5 kV, 1.62 nA, 102 K). The data were fitted with four Lorentzians. The emission lines at 206 nm and 208 nm are strongly reduced when using AlBr$_3$.

An altered incorporation mechanism, placing defects/carbon on interstitial lattice sites or the formation of defect/carbon-related complexes that are not excited by CL, needs to be considered when interpreting the CL results.

Due to the intrinsically weak photoluminescence of AlN, we could not clearly distinguish the difference between the layer deposited with TMAl and that deposited with AlBr$_3$ by PL.

As for the HRXRD ω-scan FWHM for this sample, it was 386 arcsec for the 0002 reflection, and 709 for the -1102 reflection, which are both similar to the quality of the underlying TMAl layer deposited with TMAl (**Figure S4**). We performed sheet resistance by Eddy-current measurements. As expected for AlN, the layer was highly insulating with values above 100 kΩ sq$^{-1}$, clearly indicating that the layer deposited using AlBr$_3$ was highly resistive.

## 4. Conclusion

In this work, we have demonstrated the epitaxial growth of group 13-nitrides in a commercial reactor by using C-free precursors. We have identified brominated precursors as the most promising C-free precursors due to their lower corrosiveness and reasonable vapor pressure, which enables the achievement of reasonable molar flows with a proper vaporizer setup. We described the growth of GaN layers by using GaBr$_3$ and AlN layers by using AlBr$_3$. The optical properties measured by PL and CL indicate a clear reduction in the electrically and optically active carbon-related defects in the layers, consistent with the use of carbon-free precursors. While the precursors' purity needs to be improved by the manufacturers, further exploration of the optimum operating window is in progress and will allow us to push towards higher material quality, higher growth rates, and reduced contamination. These initial results demonstrate the feasibility of this process and the high likelihood that GaN, AlN, and probably AlGaN alloys can be deposited with the Br-based CVD process.

**Acknowledgements**

The authors would like to acknowledge S. Mueller, C. Manz, M. Prescher, N. Brueckner, P. Straňák, R. Aidam, H. Mueller, P. Neininger, as well as all the colleagues at Fraunhofer IAF who have helped and supported these activities. Isabel Streicher is acknowledged for her former support on the CVD process with low-vapor-pressure sources and for reviewing the paper.
C. Bizzarri is acknowledged for the support in developing the conceptualization of the chemistry approach and reviewing the paper.
Dockweiler Chemicals is acknowledged for the relevant discussions about halogenated Al- and Ga-precursors, as well as for supporting the supply and usage of GaBr$_3$.
The authors declare no conflict of interest.

**Data Availability Statement**

The data that support the findings of this study are available from the corresponding author upon reasonable request.

# References

[1] K. Ohnishi, S. Kawasaki, N. Fujimoto, et al., "Vertical GaN p[+]-n junction diode with ideal avalanche capability grown by halide vapor phase epitaxy", *Appl. Phys. Lett.* 119, no. 15 (2021): 152102.

[2] S. Mukhopadhyay, P. Seshadri, M. Haque, et al., "Demonstration of Si-doped Al-rich regrown Al(Ga)N films on AlN/sapphire with >1015/cm3 carrier concentration using CCS-MOCVD reactor", *Appl. Phys. Lett.* 125, no. 22 (2024): 222102.

[3] T. Ciarkowski, N. Allen, E. Carlson, et al., "Connection between Carbon Incorporation and Growth Rate for GaN Epitaxial Layers Prepared by OMVPE". *Materials* 12 (2019): 2455

[4] M. Kim, A. Papamichail, D. Q. Tran, et al., "Thin-channel AlGaN/GaN/AlN double heterostructure HEMTs on AlN substrates via hot-wall MOCVD", *Applied Physics Letters* 127 (2025): 032104.

[5] S. Kaneki, T. Konno, H. Mori and H. Fujikura, "Quartz-free hydride vapor phase epitaxy for production of large size GaN-on-GaN epitaxial wafers" *Appl. Phys. Express* 18 (2025): 055502.

[6] T. Narita, K. Tomita, Y. Tokuda, T. Kogiso, M. Horita, T. Kachi, "The origin of carbon-related carrier compensation in p-type GaN layers grown by MOVPE", *J. Appl. Phys.* 124, no. 21 (2018): 215701.

[7] I. Bryan, Z. Bryan, S. Washiyama, et al., "Doping and compensation in Al-rich AlGaN grown on single crystal AlN and sapphire by MOCVD", *Appl. Phys. Lett.* 112 (2018): 062102.

[8] D. Wickramaratne, M. Siford, M. Mollik, J. L. Lyons, and M. E. Zvanut, "Direct evidence for carbon incorporation on the nitrogen site in AlN", *Phys. Rev. Materials* 8 (2024): 094602.

[9] N. Zagni, A. Chini, F. M. Puglisi, P. Pavan, G. Verzellesi, "On the Modeling of the Donor/Acceptor Compensation Ratio in Carbon-Doped GaN to Univocally Reproduce Breakdown Voltage and Current Collapse in Lateral GaN Power HEMTs", *Micromachines* 12, no. 6 (2021): 709.

[10] Y. Zhang, V. G. T. Vangipuram, K. Zhang, H. Zhao, "Investigation of carbon incorporation in laser-assisted MOCVD of GaN", *Appl. Phys. Lett.* 122, no. 16 (2023): 162101

[11] D. Crippa, "*Silicon Epitaxy*", Book in Semiconductor and Semimetals 72 (2001), ISBN: 978-0-12-752181-7, ISSN: 0080-8784

[12] S. Leone, O. Kordina, A. Henry, S Nishizawa, O. Danielsson, E. Janzén, "Gas-phase modeling of chlorine-based chemical vapor deposition of silicon carbide", *Crystal growth & design* 12, no. 4 (2012): 1977.

[13] S. Leone, F. Beyer, H. Pedersen, O. Kordina, A. Henry, E. Janzén, "High growth rate of 4H-SiC epilayers on on-axis substrates with different chlorinated precursors", *Crystal growth & design* 10, no. 12 (2010): 5334.

[14] S. Leone, A. Henry, E. Janzén, S. Nishizawa, "Epitaxial growth of SiC with chlorinated precursors on different off-angle substrates", *Journal of Crystal Growth* 362 (2013): 170.

[15] F. La Via, G. Galvagno, F. Roccaforte, et al., "High growth rate process in a SiC horizontal CVD reactor using HCl", *Microelectronic engineering* 83, no. 1 (2006): 48.

[16] S. Leone, F. Beyer, A. Henry, O. Kordina, E. Janzén, "Chloride-based CVD of 3C-SiC epitaxial layers on 6H (0001) SiC", *Physica Status Solidi (RRL)* 4, no. 11 (2010): 305.

[17] L. Calcagno, G. Izzo, G. Litrico, et al., "Optical and electrical properties of 4H-SiC epitaxial layer grown with HCl addition", *Journal of Applied Physics* 102, no. 4 (2007): 043523.

[18] S. Leone, F. Beyer, H. Pedersen, O. Kordina, A. Henry, E. Janzén, "Growth of smooth 4H-SiC epilayers on 4° off-axis substrates with chloride-based CVD at very high growth rate", *Materials research bulletin* 46, no. 8 (2011): 1272.

[19] S. Leone, F. Beyer, A. Henry, C. Hemmingsson, O. Kordina, E. Janzén, "Chloride-based SiC epitaxial growth toward low temperature bulk growth", *Crystal growth & design* 10, no. 8 (2010): 3743.

[20] M. Yazdanfar, O. Danielsson, E. Kalered, et al., "Brominated Chemistry for Chemical Vapor Deposition of Electronic Grade SiC", *Chemistry of Materials* 27, no. 3 (2015): 793.

[21] T. L Chu, "Gallium nitride films", *J. Electrochem. Soc.* 118 (1971): 1200.

[22] J. Chen, U. Forsberg, E. Janzén, "Impact of residual carbon on two-dimensional electron gas properties in $Al_xGa_{1-x}N$/GaN heterostructure", *Applied Physics Letters* 102 (2013): 193506.

[23] H. Yacoub, C. Mauder, S. Leone, et al., "Effect of Different Carbon Doping Techniques on the Dynamic Properties of GaN-on-Si Buffers", *IEEE Transactions on Electron Devices* 64, no. 3 (2017): 991.

[24] A. K. Dhasiyan, F. W. Amalraj, S. Jayaprasad, et al., "Epitaxial growth of high-quality GaN with a high growth rate at low temperatures by radical-enhanced metalorganic chemical vapor deposition", *Science Reports* 14 (2024): 10861.

[25] https://www.aixtron.com/en, A. Debald "Optimization of the epitaxial drift region of GaN-on-GaN vertical devices", at the *Internation Wokshop of Nitrides* (2024).

[26] J. Singhal, J. Encomendero, Y. Cho, et al.; "Molecular beam homoepitaxy of N-polar AlN on bulk AlN substrates", *AIP Advances* 12, no. 9 (2022): 095314.

[27] Y. Kumagai, K. Goto, T. Nagashima, R. Yamamoto, M. Boćkowski, J. Kotani, *Appl. Phys. Express* 15 (2022): 115501.

[28] J.H. Leach, K. Udwary, G. Dodson, T.B. Tran, H.A. Splawn, "Low-Pressure, Modified Halide Vapor-Phase Epitaxy for Chemically Pure GaN Epilayers", *Phys. Status Solidi B* 261 (2024): 2400035.

[29] T. Hirasaki, K. Hanaoka, H. Murakami, Y. Kumagai, A. Koukitu, "Thermodynamic analysis of InGaN-HVPE growth using group-III chlorides, bromides, and iodides", *Phys. Status Solidi C* 10 (2013): 413.

[30] S. Leone, C. Manz, H. Menner, J. Wiegert, J. Ligl, "Method and apparatus for manufacturing a semiconductor layer and substrate provided therewith". Patent ID: US 11996287.

[31] S. Leone, J. Ligl, C. Manz, L. Kirste, T. Fuchs, H. Menner, M. Prescher, J. Wiegert, A. Zukauskaite, R. Quay, O. Ambacher, "Metal-Organic Chemical Vapor Deposition of Aluminum Scandium Nitride", *physica status solidi (RRL) – Rapid Research Letters* 14, no. 1 (2020): 1900535.

[32] I. Streicher, S. Leone, L. Kirste, C. Manz, P. Straňák, M. Prescher, P. Waltereit, M. Mikulla, R. Quay, O. Ambacher, "Enhanced AlScN/GaN heterostructures grown with a novel precursor by metal–organic chemical vapor deposition", *physica status solidi (RRL) – Rapid Research Letters* 17, no. 2 (2022): 2200387.

[33] I. Streicher, S. Leone, C. Manz, L. Kirste, M. Prescher, P. Waltereit, M. Mikulla, R. Quay, O. Ambacher, "Effect of AlN and AlGaN interlayers on AlScN/GaN heterostructures grown by metal–organic chemical vapor deposition", *Crystal Growth & Design* 23, no. 2 (2023): 782.

[34] I. Streicher, S. Leone, M. Zhang, T. S. Tlemcani, M. Bah, P. Straňák, L. Kirste, M. Prescher, A. Yassine, D. Alquier, O. Ambacher, "Understanding Interfaces in AlScN GaN Heterostructures", *Advanced Functional Materials* 34, no. 39 (2024): 2403027.

[35] S. Krause, I. Streicher, P. Waltereit, L. Kirste, P. Bruckner, S. Leone, "AlScN/GaN HEMTs grown by metal-organic chemical vapor deposition with 8.4 W/mm output power and 48% power-added efficiency at 30 GHz", *IEEE Electron Device Letters* 44, no. 1 (2023): 17.

[36] C. Manz, S. Leone, L. Kirste, J. Ligl, K. Frei, T. Fuchs, M. Prescher, P. Waltereit, M. A. Verheijen, A. Graff, M. Simon-Najasek, F. Altmann, M. Fiederle, O. Ambacher, "Improved AlScN/GaN heterostructures grown by metal-organic chemical vapor deposition", *Semiconductor Science and Technology 36*, no. 3 (2021): 034003.

[37] N. Wolff, G. Schoenweger, I. Streicher, M. R. Islam, N. Braun, P. Straňák, L. Kirste, M. Prescher, A. Lotnyk, H. Kohlstedt, S. Leone, L. Kienle, S. Fichtner, "Demonstration and STEM

Analysis of Ferroelectric Switching in MOCVD-Grown Single Crystalline Al$_{0.85}$Sc$_{0.15}$N", *Advanced Physics Research* 3, no. 5 (2024): 2300113.

[38] N Wolff, T Grieb, G Schönweger, et al., "Electric field-induced domain structures in ferroelectric AlScN thin films", *J. Appl. Phys.* 137 (2025): 084101.

[39] S. Leone, I. Streicher, M. Prescher, P. Straňák, L. Kirste, "Metal-Organic Chemical Vapor Deposition of Aluminum Yttrium Nitride", *physica status solidi (RRL) – Rapid Research Letters* 17, no. 10 (2023): 2300091.

[40] I. Streicher, P. Straňák, L. Kirste, M. Prescher, S. Mueller, S. Leone, "Two-dimensional electron gases in AlYN/GaN heterostructures grown by metal–organic chemical vapor deposition ", *APL Materials* 12, no. 5 (2024)**:** 051109.

[41] J. Ligl, S. Leone, C. Manz, L. Kirste, P. Doering, T. Fuchs, M. Prescher, O. Ambacher, "Metalorganic chemical vapor phase deposition of AlScN/GaN heterostructures", *Journal of Applied Physics* 127, no. 19 (2020): 195704.

[42] O. Ambacher, B. Christian, M. Yassine, M. Baeumler, S. Leone, R. Quay, "Polarization induced interface and electron sheet charges of pseudomorphic ScAlN/GaN, GaAlN/GaN, InAlN/GaN, and InAlN/InN heterostructures", *Journal of Applied Physics* 129, no. 20 (2021): 204501.

[43] V Stanishev, I Streicher, A Papamichail, et al., "2DEG properties of AlScN/GaN and AlYN/GaN HEMTs determined by terahertz optical Hall effect", *Frontiers in Electronic Materials* 5 (2025), 1622176.

[44] K. Nomoto, I. Streicher, TS. Nguyen, et al., "MOCVD AlYN/GaN HEMTs with 66.5 mV/decade sub-threshold swing and 109 on/off ratio", *Applied Physics Letters* 126, no. 22 (2025): 223509.

[45] H. Lu, G. Schönweger, N. Wolff, Z. Ding, et al., "Al$_{1-x}$Sc$_x$N-Based Ferroelectric Domain-Wall Memristors", *Advanced Functional Materials* (2025): 2503143.

[46] A. Yassine, M. Yassine, I. Streicher, et al., "Electrical properties of SiN$_y$/Al$_{1-x}$Sc$_x$N/GaN-based metal–insulator–semiconductor structures", *Journal of Applied Physics* 137, no. 19 (2025): 194501.

[47] N. N. Greenwood, A. Earnshaw, *Chemistry of the elements,* Elsevier Ltd. (2010), ISBN 978-0-7506-3365-9.

[48] B. Brunetti, V. Piacente, P. Scardala, "Vapor Pressures of Aluminum Tribromide and Aluminum Triiodide and Their Standard Sublimation Enthalpies", *J. Chem. Eng. Data* 55 (2010): 2164.

[49] B. Brunetti, V. Piacente, P. Scardala, "A Study on the Sublimation of Gallium Tribromide", *J. Chem. Eng. Data* 54 (2009): 2273.

[50] German patent application "Schicht und Verfahren zu ihrer Herstellung" 10 2025 132 956.7, submitted on 18.08.2025.

[51] T. Coenen, "Monte Carlo Beam Tracing Simulations for Incoherent Cathodoluminescence Emission", Delmic Technical Note (2019).

[52] D. Drouin, A.R. Couture, D. Joly et al., "CASINO V2.42—A Fast and Easy-to-use Modeling Tool for Scanning Electron Microscopy and Microanalysis Users", *Scanning*, 29 (2007): 92.

[53] H. Demers, N. Poirier-Demers, A.R. Couture, D. Joly et al., "Three-dimensional electron microscopy simulation with the CASINO Monte Carlo software", *Scanning* 33 (2011): 135.

[54] B. Gil, O. Briot, "Internal structure and oscillator strengths of excitons in strained α-GaN." , *Physical Review B* 55, no.4 (1997): 2530.

[55] R. Khan, K. Narang, A. Arora et al., "Improvement in the crystalline quality of GaN and defects analysis using cathodoluminescence". *Materials Today Proceedings* vol. 36, part 3 (2021): 631.

[56] Y. Xu, X. Yang, P. Zhang, X. Cao, "Influence of intrinsic or extrinsic doping on lattice locations of carbon in semi-insulating GaN.", *Applied Physics Express* 12, no. 6 (2019): 061002.

[57] J. L. Lyons, A. Janotti, C. G. Van de Walle, "Carbon impurities and the yellow luminescence in GaN.", *Applied Physics Letters* 97 (2010): 152108.

[58] S. G. Christenson, W. Xie, Y. Y. Sun, S. B. Zhang, "Carbon as a source for yellow luminescence in GaN: Isolated CN defect or its complexes.", *Journal of Applied Physics* 118, no. 13 (2015): 135708.

[59] M. A. Reshchikov, Y. T. Moon, X. Gu et al. "Unstable luminescence in GaN and ZnO." *Physica B: Condensed Matter* 376 (2006): 715.

[60] D. O. Demchenko, I. C. Diallo, and M. A. Reshchikov, "Hydrogen-carbon complexes and the blue luminescence band in GaN.", *Journal of Applied Physics* 119 (2016): 035702.

[61] M. A. Reshchikov, M. Vorobiov, D. O. Demchenko et al., "Two charge states of the $C_N$ acceptor in GaN: Evidence from photoluminescence", *Phys. Rev. B* 98, (2018): 125207.

[62] M. A. Reshchikov, "Fine Structure of the Carbon-Related Blue Luminescence Band in GaN", *Solids* 3, no.2 (2022): 231.

[63] M. A. Reshchikov, O. Andrieiev, M. Vorobiov et al., "Stability of the $C_NH_i$ Complex and the Blue Luminescence Band in GaN", *Phys. Status Solidi B* 258 (2021): 2100392.

[64] L. Kirste, T. Lim, R. Aidam, S. Müller, P. Waltereit, O. Ambacher, "Structural properties of MBE AlInN and AlGaInN barrier layers for GaN-HEMT structures", *Phys. Status Solidi A* 207 (2010): 1338.

[65] Y. G. Cao, X. Chen, Y. Lan et al., "Blue emission and Raman scattering spectrum from AlN nanocrystalline powders.", *Journal of Crystal Growth* 213, no. 1-2 (2000): 198.

[66] S. Tungasmita, P.O.A. Persson, L. Hultman, J. Birch, "Pulsed low-energy ion-assisted growth of epitaxial aluminum nitride layer on 6*H*-silicon carbide by reactive magnetron sputtering", *J. Appl. Phys.* 91 (2002): 3551.

[67] N. Teofilov, K. Thonke, R. Sauer et al., "Near band-edge transitions in AlN thin films grown on different substrates", *Diamond and Related Materials* 10 (2001): 1300.

[68] M. Feneberg, B. Neuschl, K. Thonke et al., "Sharp bound and free exciton lines from homoepitaxial AlN.", *physica status solidi (a)* 208, no. 7 (2011): 1520.

[69] L. van Deurzen, J. Singhal, J. Encomendero et al., "Excitonic and deep-level emission from N- and Al-polar homoepitaxial AlN grown by molecular beam epitaxy", *APL Mater.* 11, no. 8 (2023): 081109.

**Table of contents**

## Chemical Vapor Deposition of Nitrides by Carbon-free Brominated Precursors

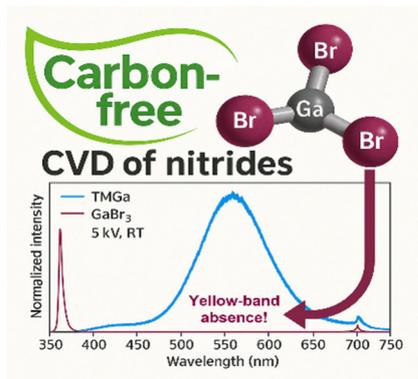

Brominated precursors $GaBr_3$ and $AlBr_3$ enable the growth of epitaxial layers of (Al,Ga)N by chemical vapor deposition with carbon-free precursors. The GaN epitaxial layers do not show any yellow-band luminescence according to cathodoluminescence measurements, unlike GaN layers grown using the standard $Ga(CH_3)_3$ (TMGa) precursor.

Supporting Information

## Chemical Vapor Deposition of Nitrides by Carbon-free Brominated Precursors

*Stefano Leone\*, Teresa Duarte, Hanspeter Menner, Jannik Richter, Lutz Kirste, Sven Maegdefessel, Felix Hoffmann, Byeongchan So, and Ruediger Quay*

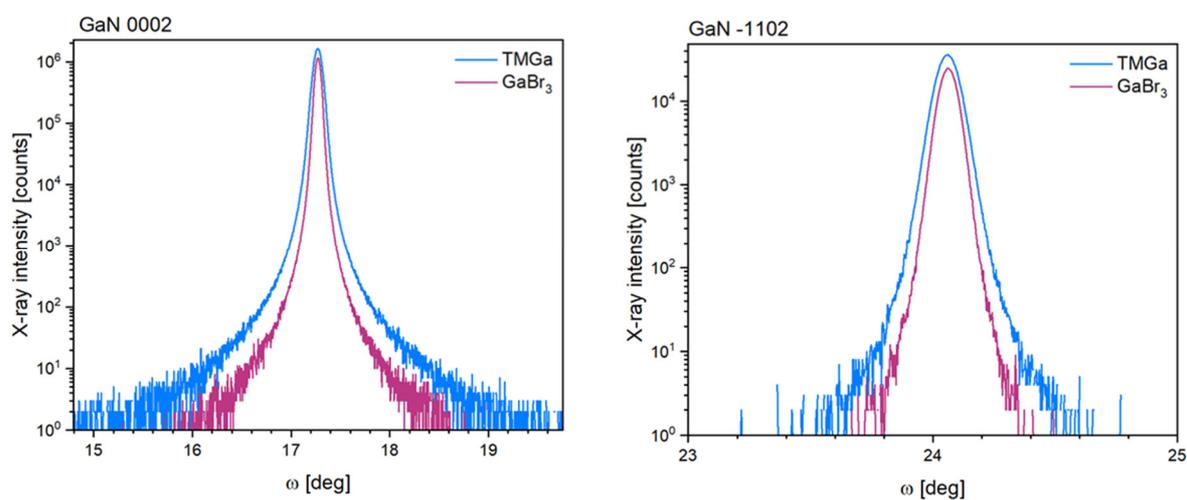

**Figure S1**: HRXRD ω-scan FWHM of the 2 μm GaN epitaxial layer deposited with TMGa as Ga-precursor (blue), and the 300 nm GaN layer deposited with $GaBr_3$ as Ga-precursor (purple) on top of the GaN template.

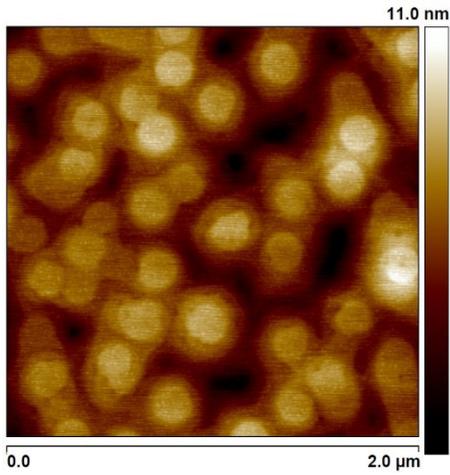
**Figure S2**: AFM analysis in tapping mode on a 300 nm thick AlN epitaxial layer deposited on AlN/Al$_2$O$_3$ templates with AlBr$_3$. The RMS is 1.2 nm on a 2x2 µm$^2$ scan area.

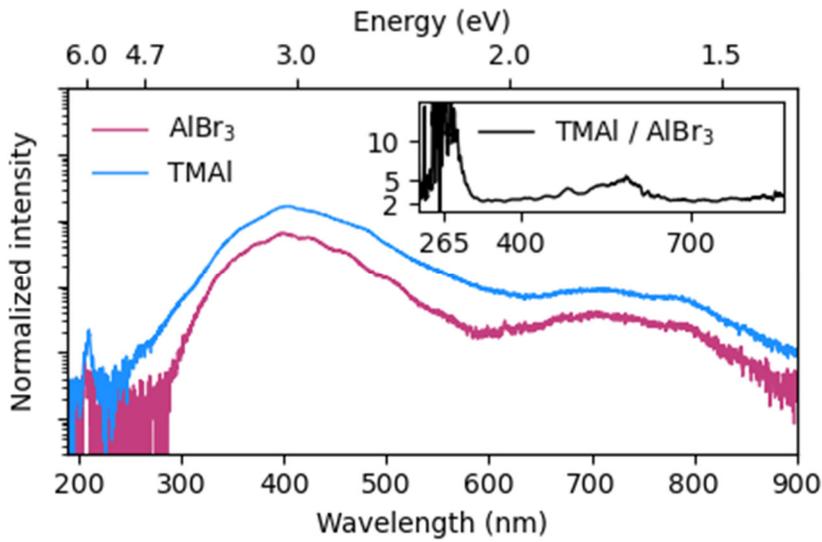
**Figure S3**: CL spectroscopy of AlN epitaxial layers grown with AlBr$_3$ (purple) and TMAl (blue), normalized to the free-exciton emission at 204.5 nm (5 kV, 1.62 nA, 102 K). The inset shows the ratio of both spectra.

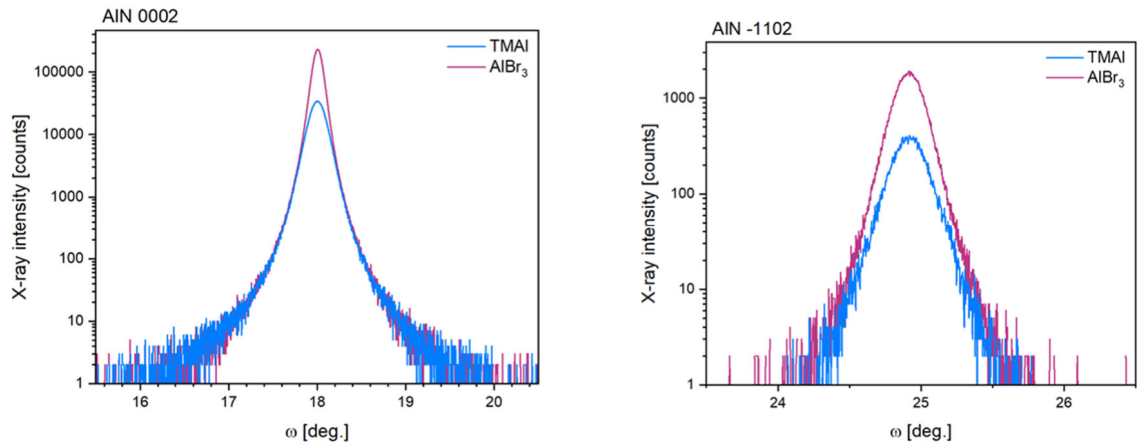

**Figure S4**: HRXRD ω-scan FWHM of the 1 μm thick AlN epitaxial layer deposited with TMAl as Al-precursor (blue), and the 300 nm AlN layer deposited with AlBr$_3$ as Al-precursor (purple) on top of the AlN template.